\documentclass[12pt,preprint]{aastex}
\usepackage{amsmath}
\usepackage{natbib}

\begin{document}

\bibliographystyle{apj}

\title{The Nature of the Atmosphere of the Transiting Super-Earth GJ 1214b}

\author{Eliza Miller-Ricci}

\affil{Department of Astronomy and Astrophysics, University of California, 
      Santa Cruz, CA 95064}

\email{elizamr@ucolick.org}

\author{Jonathan J. Fortney}

\affil{Department of Astronomy and Astrophysics, University of California, 
      Santa Cruz, CA 95064}

\begin{abstract}

The newly discovered planet GJ 1214b is the first known transiting super-Earth
requiring a significant atmosphere to explain its observed mass and radius.
Models for the structure of this planet predict that it likely possesses a
H-He envelope of at least 0.05\% of the total mass of the planet.  However,
models without a significant H-He atmosphere are not entirely ruled out by the
available data.  Here we explore a range of possible atmospheres for the 
planet, ranging from solar composition gas, to pure CO$_2$ or water (steam).  
We present transmission and emission spectra for each of these cases.  We find
that, if GJ 1214b possesses a hydrogen-rich atmosphere as expected, then the
primary transit depth for such an atmosphere would vary at a level of
up to 0.3\% as a function of wavelength, relative to the background light
of its M-dwarf host star.  Observations at this level of precision are 
potentially obtainable with current space-based instrumentation.  Successful 
detection of the transmission signature of this planet at the $\sim$0.1\% 
level would therefore provide 
confirmation of the hydrogen-rich nature of the planetary atmosphere. It
follows that transmission spectroscopy at this level of precision
could provide a first glimpse into answering the question of whether planets in
the super-Earth mass regime (1 - 10 M$_{\earth}$) more closely resemble large 
terrestrial planets or small gas giant planets.

\end{abstract}


\section{Introduction}

The newly discovered class of super-Earths (planets with masses between
1 and 10 M$_{\earth}$) brings planet detection into a mass regime that has not
previously been studied, due to the lack of such planets in our Solar 
System.  The recent detection of the first \textit{transiting} super-Earths 
\citep{leg09, cha09} has allowed for the first 
measurements of mass \textit{and} radius, and 
therefore bulk density of these objects.  Since super-Earths lie in an 
intermediate mass range between terrestrial planets and the gas/ice giants
of our Solar System, the question arises as to where the dividing line lies
between these two classes of planets.  Super-Earths with thick hydrogen/helium
atmospheres (``sub-Neptunes'') may have experienced formation histories that 
are similar to those of other gas giants, whereas predominantly solid planets 
that are true 
scaled up analogs of the terrestrial planets in our Solar System would instead 
be thought of as ``super'' or large Earths.  


GJ 1214b \citep{cha09} is the second known transiting super-Earth, and also the
first such planet to likely posess significant gaseous envelope.  The planet's
mass (6.6 M$_{\earth}$) and radius (2.7 R$_{\earth}$) imply a density that
is too low to be explained by rock and water ice alone.  Indeed, a comparison 
against theoretical mass-radius relationships predicted for solid exoplanets 
\citep{val06, for07, sea07, sot07} reveals that GJ1214b lies in a region of 
parameter space where a significant H/He atmosphere is the most likely 
explaination for the planet's observed properties.  In their discovery paper, 
\citet{cha09} state that a low-density composition for the planet, made up of 
75\% water (ice), 22\% silicate rock, and 3\% iron would still necessitate a 
H/He atmosphere of 0.05 mass percent to explain the planet's radius.  This 
corresponds to a H/He atmosphere of 
approximately 70 bars.  Such a planet lies somewhere on the spectrum between
terrestrial planets, and gas or ice giants, making this a highly interesting
new object for study.  

The presence of a significant H/He atmosphere on GJ 1214b is not confirmed
however by the mass and radius measurements for this planet.  \citet{rog10} 
find that the observed mass and radius of GJ 1214b can be explained if the 
planet is composed of at least 47\% water accompanied by a massive 
\textit{steam} atmosphere.  Additionally, \citet{cha09} 
point out that their measured 
radius of the planet's M-dwarf host star, GJ 1214, is inconsistent with 
theoretical models.  If in fact the radius of the star is 15\% smaller, as 
predicted by models \citep{bar98}, then the radius of the planet,
GJ 1214b, would also be smaller by a comparable amount -- resulting in a planet
that does not require a large atmosphere to explain its observed radius.  


Given the uncertainties in the nature of the atmosphere of GJ 1214b, follow-up
observations aimed at detecting the spectral signature of the planet's 
atmosphere will be necessary to confirm its molecular makeup.  This could be 
accomplished via secondary eclipse (emission) or occultation (transmission) 
spectroscopy.  Both methods have been shown to be successful in characterizing 
the atmospheres of gas giant exoplanets \citep[e.g.][]{cha02, dem05, swa08, 
knu08}.

\citet{mil09} have pointed out that transmission 
spectroscopy is an excellent tool for determining the hydrogen content of
an exoplanet atmosphere.  Since the depth of spectral features in transmission
is proportional to the atmospheric scale height (which in turn scales
inversely proportionally to the mean molecular weight of the planetary 
atmosphere), a hydrogen-rich atmosphere will exhibit spectral features in 
transmission that are up to a factor of 20 larger than those of a 
hydrogen-depleted CO$_2$ 
atmosphere.  For GJ 1214b, the expected signature of a
hydrogen-rich atmosphere would manifest itself on a level of $\sim$0.1\% 
relative to the M-dwarf (M4.5) host 
star.  Such precision has been shown to be obtainable with current space-based 
observatories -- both Hubble \citep[e.g.][]{swa08, pon09} and Spitzer 
\citep[e.g.][]{dem07}.  In this paper we present 
calculations of transmission and emission spectra for possible atmospheres on 
the planet GJ 1214b, ranging in composition from hydrogen-rich to 
hydrogen-poor.  We present our model
and its parameters in Section~\ref{methods}.  We then present our resulting
spectra in Section~\ref{results}, and we conclude with some thoughts on 
combining our models with observations to constrain the nature of GJ 1214b's 
atmosphere in Section~\ref{conclusion}.

\section{Methodology \label{methods}}

\subsection{Range of Model Parameters}

We consider atmospheres for GJ 1214b ranging from hydrogen-dominated solar 
composition gas to hydrogen- and helium-depleted atmospheres made up of 
predominantly heavier molecules. 
The six different cases of atmospheric composition that we investigate are:

\begin{enumerate}
\item \textit{Solar composition atmosphere} -- For this case and the two that 
follow, the molecular species that make up the atmosphere appear in 
equilibrium abundances (see Section~\ref{model}, for more details).
\item \textit{30 $\times$ Enhanced metallicity atmosphere} -- Here we
use the base composition of atmosphere \# 1, but in this case the abundances of
all species except for H and He are enhanced by a factor of 30.  
\item \textit{50 $\times$ Enhanced metallicity atmosphere} -- Here the 
metallicity is enhanced 50 times relative to solar.
\item \textit{100\% water (steam) atmosphere}
\item \textit{50\% water, 50\% CO$_2$ atmosphere}
\item \textit{CO$_2$ atmosphere plus trace gasses} -- This model atmosphere
is composed of 96.5\% CO$_2$ with other trace gasses appearing in Venusian 
abundances, notably 3.5\% N$_2$ and 20 ppm H$_2$O.
\end{enumerate}

Simple energy balance arguments for the atmosphere of GJ1214b yield a 
day-side equilibrium temperature of 660 K for redistribution of incident energy
over the day side only, and 555 K for efficient redistribution over the entire 
planet, if the planet's Bond albedo is zero.  \citet{mar99} found that all 
planets orbiting M-dwarfs should have very low albedos, in the absence of 
clouds or surfaces that efficiently scatter incoming starlight. The reason for 
this is that M-stars give off most of their light in the IR, rather than the 
visible, and at these wavelengths Rayleigh scattering is extremely inefficient.
The planet therefore absorbs nearly all of the incident starlight, rather than 
scattering it away.  For H/He dominated atmospheres we calculate 
self-consistent models that justify this zero-albedo assumption.  To bracket 
the likely range of atmospheric temperatures, we compute models with day-side 
and planet-wide energy redistribution -- $T_{eq} = 660$ and 555 K, 
respectively.

\subsection{Description of the Model Atmosphere \label{model}}

\begin{figure}
\plotone{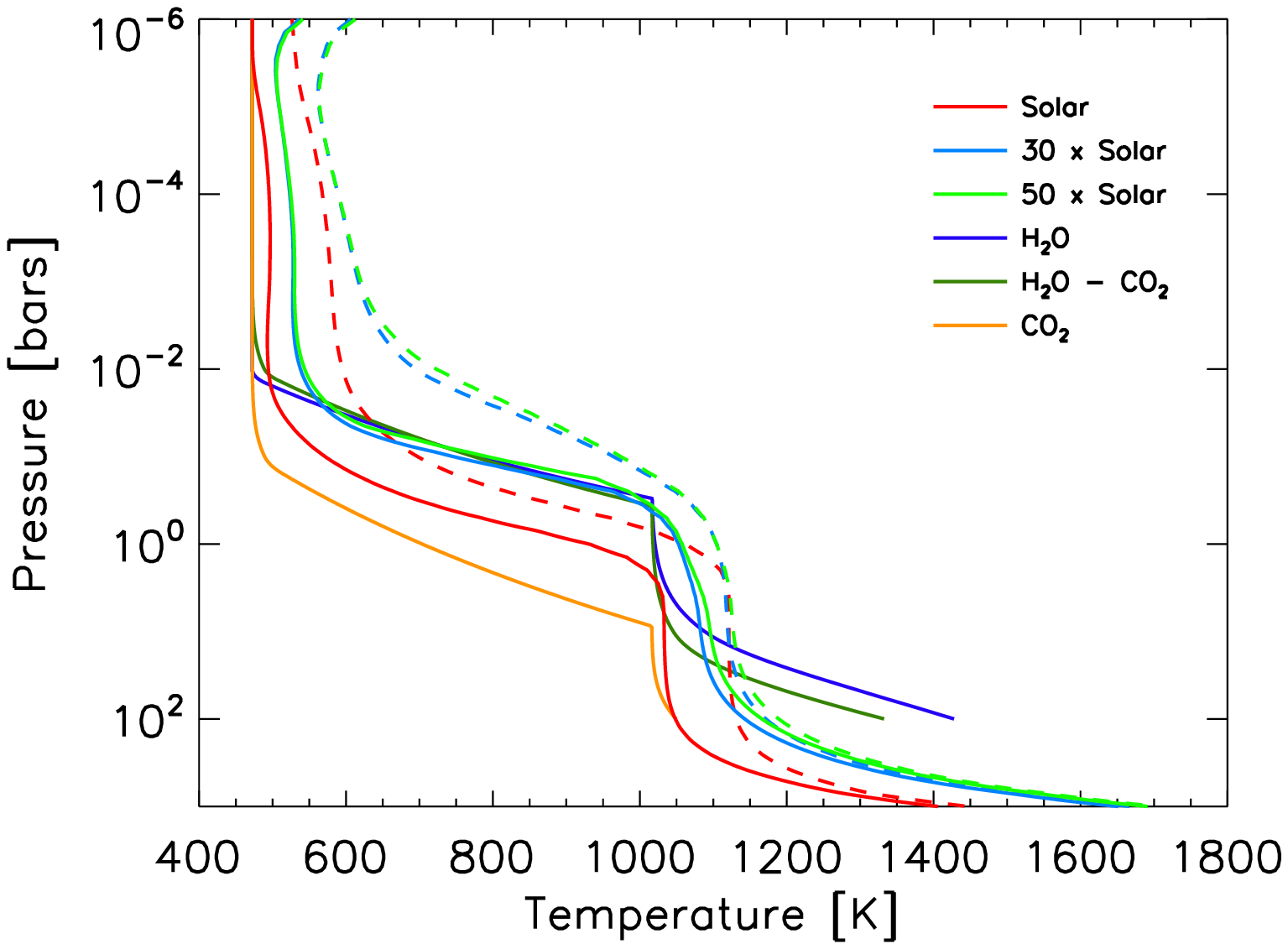}
\caption{T-P profiles for the various atmospheres presented in this paper.
        Solid lines indicate models where the day- to night-side redistribution
	of heat is efficient (T$_{eff} = 555$ K).  Dashed lines indicate models
	where the planet reradiates all of its heat from the day side of the 
	planet only (T$_{eff} = 660$ K).
        \label{t_p}}
\end{figure}

For the three hydrogen-dominated compositions that we consider (cases 1, 2, 
and 3), we calculate temperature-pressure (T-P) profiles using the LTE model 
atmosphere code of \citet{mar99b}, as applied to irradiated exoplanets by 
\citet{for05, for06, for08}.  For the remaining three hydrogen-depleted 
atmospheres (cases 4, 5, and 6), we 
calculate T-P profiles according to the model developed in \citet{mil09}. Both 
model atmospheres determine radiative equilibrium T-P profiles in hydrostatic 
equilibrium, which then switch over to adiabatic profiles in regions where the 
atmosphere is found to be convectively unstable. The Fortney et al.\ model
self-consistently calculates the atmospheric temperature structure by 
determining the levels where the incident flux from the host star is absorbed 
into the planetary atmosphere as a function of wavelength, and balancing this 
energy input against the re-emitted flux from the planet.  (See \citet{fre08} 
for a description of the opacity database.)  We find a Bond albedo of 
0.003 to 0.015 for all models, validating the approximation of a zero Bond 
albedo.  The \citet{mil09} model is simpler, in that it uses a 
gray approximation for the atmospheric opacities and temperature profile, 
following the formalism presented in \citet{han08} for an irradiated gray 
atmosphere -- their Equation 45. The resulting T-P profiles for the six 
different cases of atmospheric composition that we consider are presented
in Figure~\ref{t_p}.  Our model does not include cases with
inverted temperature profiles or sources of non-LTE emission.  Both of these
effects have been implied through observations of transiting hot-Jupiters 
\citep{swa10, rog09, mac09, knu09, knu08}.  
However, their causes are not sufficiently well-constrained to be 
self-consistently included in our model at this time.  

After determining the temperature-pressure profile, we calculate the planet's 
emitted spectrum by integrating the equation of radiative transfer through the 
planet's atmosphere.  We assume a Planckian source function, and we include the
dominant sources of molecular line opacity from 0.1 to 100 $\mu$m in the IR -- 
CH$_4$ \citep{fre08, kar94, str93}, CO , NH$_3$ \citep[][and references 
therein]{fre08}, 
CO$_2$ \citep{rot05}, and H$_2$O \citep{fre08, par97}.  For the line profiles,
we employ a Voigt broadening scheme.  In addition to the line opacities,
we also include collision-induced absorption (CIA) from H$_2$ with itself, He, 
and CH$_4$ \citep[][and references therein]{bor02, bor00}, and CO$_2$ with 
CO$_2$ interpolated between the databases of \citet{bro91} and \citet{gru97}.  
We do not include atomic line opacities (such as
the alkali metals), since we find that spectral lines from these species are 
not present at the low temperatures that are predicted in the atmosphere of GJ 
1214b.  Instead, we find that a combination of water features and Rayleigh 
scattering dominates the spectra between 0.1 and 1 $\mu$m.

For calculating the transmission spectra we use the same set of molecular
line opacities as for the emission spectra, but in this case we solve the 
transfer equation for a purely absorptive gas to determine the
line-of-sight absorption of stellar light through the atmosphere of GJ 1214b.
Here, the incident intensity from the star is calculated by 
interpolating the model M-dwarf spectra of \citet{hau99} to produce a 
spectrum at the observed temperature (3026 K) and surface gravity ($\log g = 
4.99$ in cgs units) of the host star GJ 1214 \citep{cha09}.

\begin{figure}
\begin{center}
\includegraphics[scale=0.78]{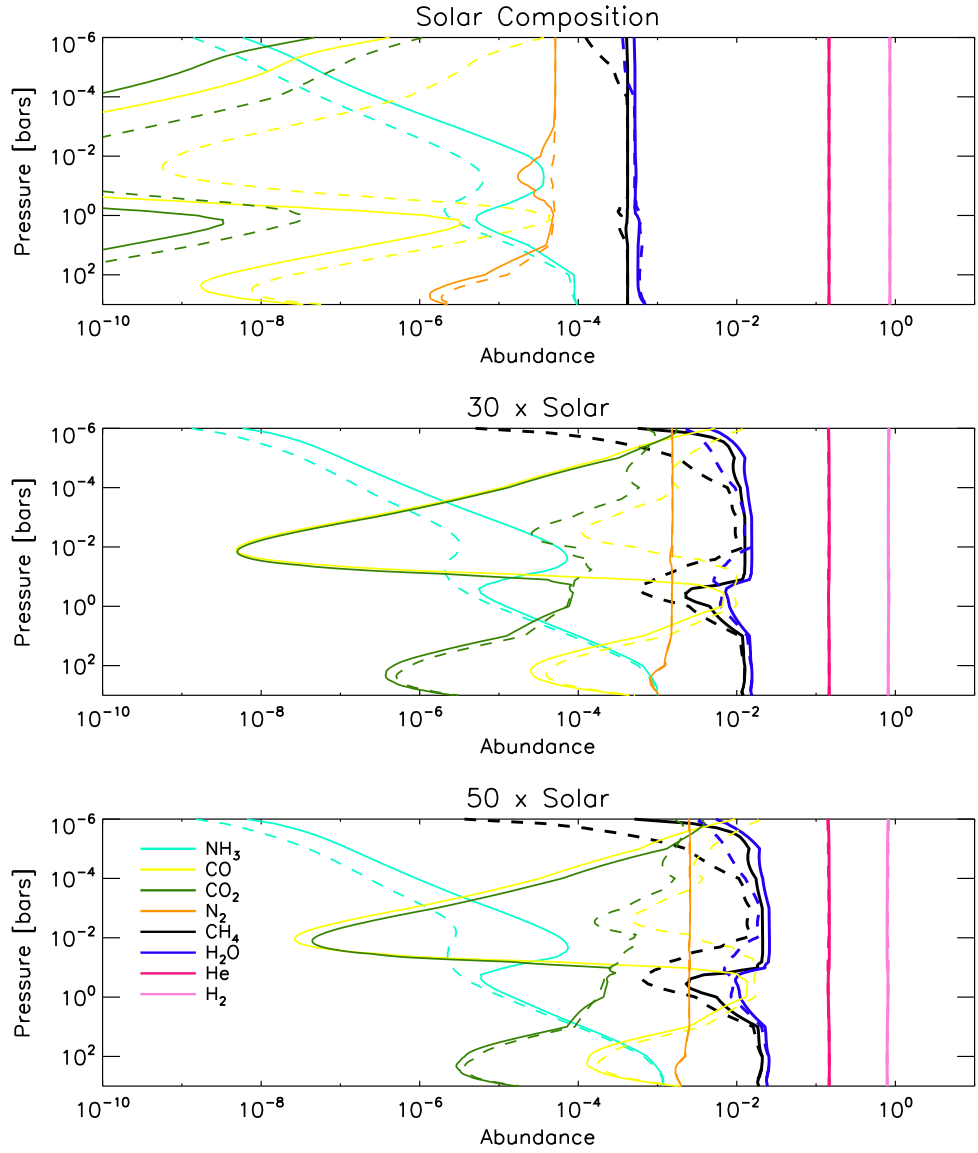}
\end{center}
\caption{Equilibrium abundances as a function of atmospheric pressure for the 
        three cases of hydrogen-dominated atmospheres.  Solid and dashed lines 
	indicate models with efficient and inefficient
	day-night heat redistribution, respectively. 
        \label{chem}}
\end{figure}

For the three hydrogen-dominated atmospheres, we determine molecular abundances
in chemical equilibrium, shown in Figure~\ref{chem}, using the code described 
in \citet{mil09}.  However, the derivation of the T-P profiles utilizes the 
equilibrium 
chemical abundances of \citet{lod02, lod06}.  Both codes produce consistent 
mixing ratios for a large number of molecular and atomic species.

\section{Probing the Composition of GJ1214b's Atmosphere \label{results}}

\begin{figure}
\begin{center}
\includegraphics[angle=90.0, scale=0.73]{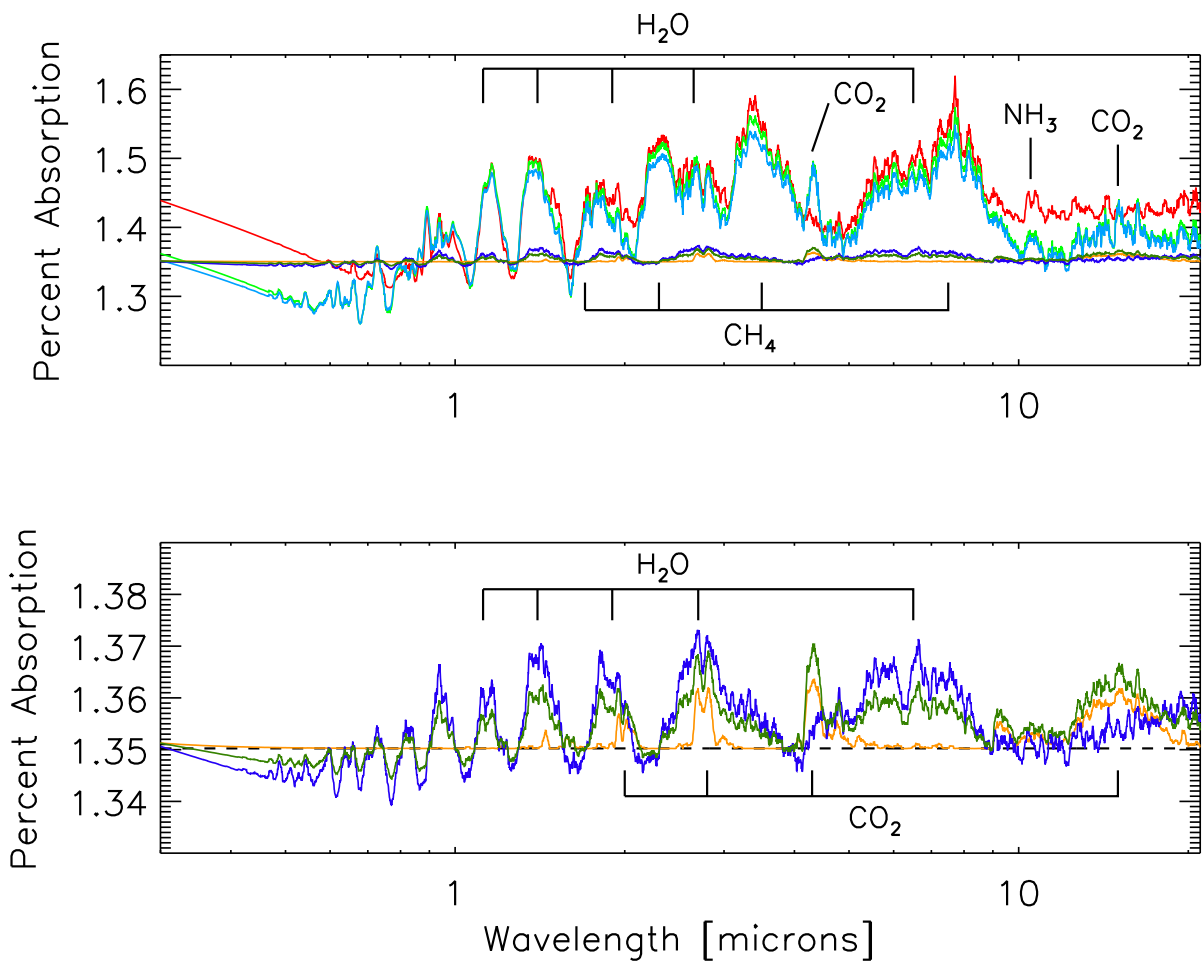}
\end{center}
\caption{Top: Transmission spectra for atmospheres with differing composition 
        --  solar composition (red), 30 $\times$ solar
	metalicity (blue), 50 $\times$ solar metalicity (light green), water
	steam (purple), 50\% water - 50\% CO$_2$ (dark green), CO$_2$ with
	trace quantities of water (orange).  All spectra are for models with
	efficient day-night circulation.  The spectra have been normalized to
	the planet's observed radius of 2.678 $R_{\earth}$ in the MEarth 
	bandpass covering 650 - 1050 microns.  Bottom: Same as above but zoomed
	in to show the spectra for the three atmospheres composed of heavier 
	molecules.  The dashed black line indicates the radius of a 
	planet with no atmosphere.
        \label{transmission}}
\end{figure}

We present our modeled transmission spectra for GJ 1214b in 
Figure~\ref{transmission}.  Assuming that the transmission
spectrum probes levels in the planetary atmosphere over a range of 10 scale
heights ($H$), then the relative change in transit depth seen throughout the 
transmission spectrum is given by:
\begin{equation}
\Delta D \sim \frac{20 H R_{pl}}{R_{*}^{2}}. \label{depth}
\end{equation}
For the case of GJ 1214b, for a hydrogen-dominated atmosphere, this results in 
a wavelength-dependent transmission spectrum that can vary in strength by up 
to 0.3\% (relative to the stellar light) as a function of 
wavelength.  Despite the small size of the transiting planet, this is a 
large transmission signature, comparable to the transmission 
signals that have been successfully observed for other (much larger) hot 
Jupiter systems. The reason for the large transmission signature is that GJ 
1214b orbits a very small
(M4.5, $R_*=0.21 R_{\odot}$) star, while the scale height for a 
hydrogen-dominated atmosphere on this planet is quite large -- approximately
150 - 200 km.  

Transmission spectra for the three hydrogen-rich atmospheres are dominated
by absorption features due to water and methane.  Methane is a particularly
interesting species in the super-Earth atmosphere, since photochemical
reactions that destroy methane may occur faster than reactions that
create this molecule \citep{zah09}.  The net result would be an 
observed methane abundance that is shifted away from its equilibrium value.  
The methane spectral features at 2.3, 3.5, and 7.5 $\mu$m are therefore 
interesting diagnostic lines for learning about photochemistry in the 
atmosphere of GJ 1214b.  Another potential consequence of methane 
destruction is the possibility of forming soots, which in principle could be an
additional gray opacity source \citep{zah09} and could furthermore cause a 
temperature inversion, therefore altering the structure of the 
planetary atmosphere.  Such effects must be modeled carefully, and are
not included in our current calculation.  Like methane, 
NH$_3$ is also highly susceptible to photodissociation, 
and the presence (or absence) of the ammonia feature at 10.5 $\mu$m is 
therefore another useful indicator of photochemical processes at work.  

The transmission spectra from the top panel of Figure~\ref{transmission} 
additionally exhibit 
CO$_2$ features in some of the hydrogen-dominated atmospheres.  
In particular, the CO$_2$ features at 4.3 and 15 $\mu$m are only present in the
atmospheres with enhanced metallicity, resulting from the higher C and O 
abundances in these atmospheres (relative to solar).  Observations at these 
particular wavelengths can therefore be used to probe the metal content of the 
super-Earth atmosphere \citep[e.g.][]{zah09b, for10}.  We caution however that 
this conclusion is dependent on our assumption 
of chemical equilibrium, which has been questioned by a number of observations
of transiting extrasolar giant planets  \citep{swa09, swa08}.

In contrast to the hydrogen-rich atmospheres, the three hydrogen-depleted 
atmospheres produce predicted transmission signatures at 
a level of 0.01 to 0.02\% relative to the M-star (bottom panel of 
Figure~\ref{transmission}), due to the reduced scaled heights in these 
atmospheres.  Observations at this level of precision will be very challenging,
given the demonstrated 
capabilities of Hubble and Spitzer.  If follows that observations with 
instruments aboard either of these telescopes would result in measurements 
that would be consistent with a flat spectrum.  Space-based observations 
revealing a flat spectrum for GJ 1214b would therefore rule out the presence of
a cloud-free hydrogen-rich atmosphere -- although H/He atmospheres with haze or
cloud layers that extend high into the atmosphere may not be excluded.  
However, spectra taken at a precision of at least several parts in 10$^{5}$ 
would be necessary to discriminate between atmospheres composed of H$_2$O, 
CO$_2$, a mix of these two gasses, or something else entirely.  Detection of 
water features between 1 and 3 $\mu$m and the CO$_2$ features at 4.3 and 15 
$\mu$m may prove to be the most useful in making this distinction.

\begin{figure}
\plotone{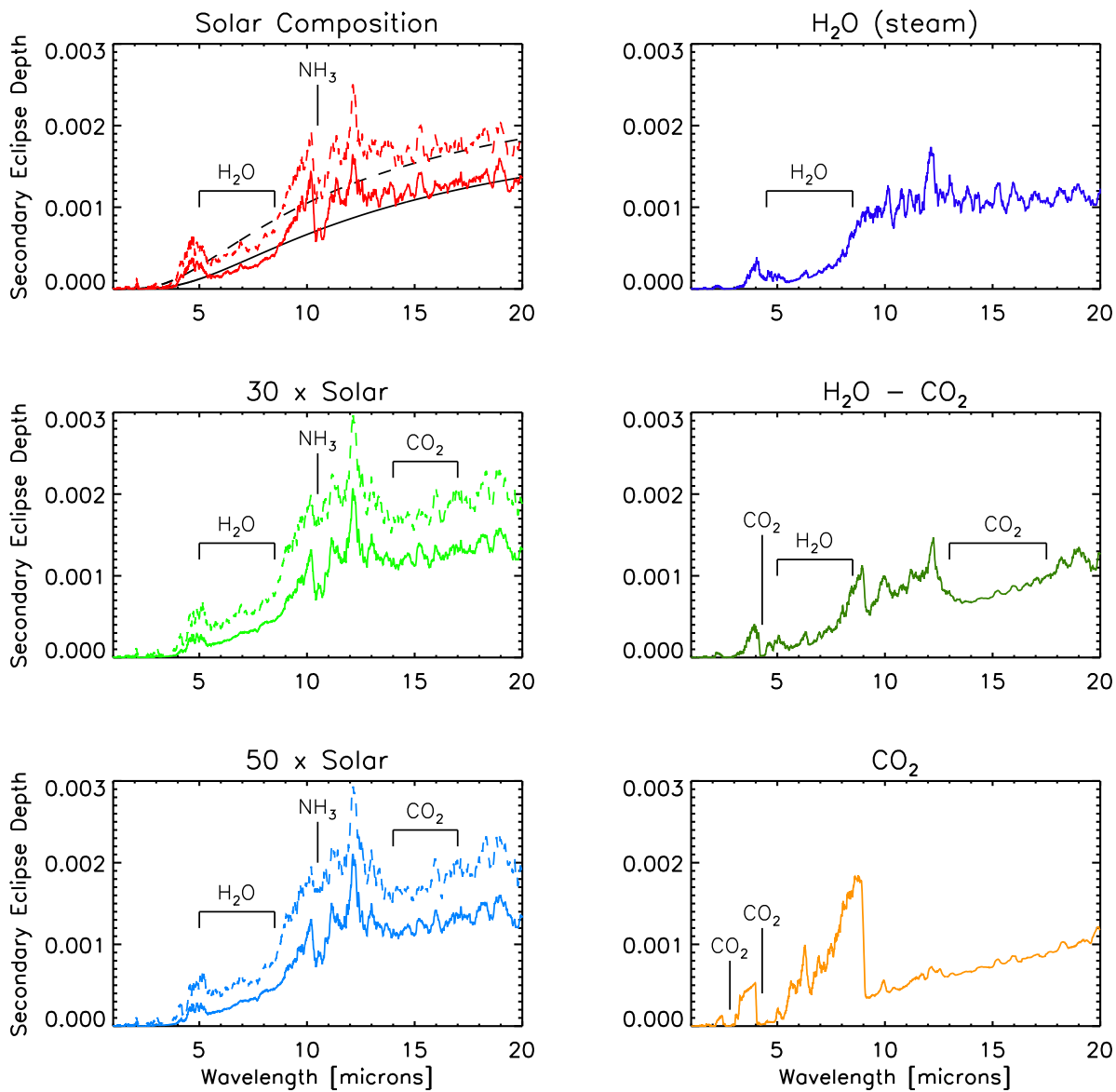}
\caption{The contrast ratio between the day-side emission from GJ 
	1214b and the emitted light from its M-dwarf host star, plotted as a 
	function of wavelength for 6 different possible atmospheric 
	compositions.  In the top left panel we overplot the 
	contrast ratios that would be expected if the planet and star both 
	emitted as blackbodies, with planetary $T_{eff}$ of 
	555 K and 660 K (thin black lines).  Dashed lines are spectra for 
	models with inefficient day-night heat redistribution.  Solid 
	lines denote models with efficient heat circulation.
        \label{emission}}
\end{figure}

We present emission spectra for GJ 1214b in Figure~\ref{emission}.  
Hydrogen-rich atmospheres with inefficient day-night 
redistribution have predicted secondary eclipse depths of up to 0.3\%.  This 
drops to approximately 0.2\% if the heat redistribution is efficient.  
These signatures are quite large and should be readily detectable with 
future instrumentation such as MIRI aboard JWST.  Unfortunately, the current 
state of high-precision space-based spectro-photometry is limited to
the wavelength range covered by the Warm Spitzer mission.  While the IRAC 3.6 
and 4.5 $\mu$m filters aboard Spitzer remain in service, all
longer wavelength bands, where the secondary eclipse depth is expected to be
the largest, are no longer operational.  Shortward of 5 $\mu$m, barring the 
presence of a 
thermal inversion or non-LTE emission from the planet's atmosphere, the 
predicted secondary eclipse depths drop below 0.1\%, which is likely to be 
prohibitively small for successful detection
with Spitzer. 

Longward of 5 $\mu$m a whole host of water, CO$_2$, methane, and ammonia 
features are predicted to be present in the secondary eclipse spectra, which 
could be used as diagnostics of atmospheric chemistry in the 
$\tau \sim 1$ region of the atmosphere.  It is interesting to note however that
a steam atmosphere and the H/He dominated atmospheres actually produce 
quite similar emission spectra, despite their very different compositions, 
since water is the dominant source of opacity in both of these cases.  
In terms of the atmospheric structure, the emission spectra 
are highly dependent on the temperature structure in the planetary atmosphere
and can therefore constrain the presence of temperature inversions in the
upper atmosphere of GJ 1214b, as well as the total amount of day-night heat
redistribution, and the possibility of non-LTE emission.  

\section{Discussion and Conclusions \label{conclusion}}

Despite its observed low density, measurements of the mass and radius of 
GJ 1214b are not sufficient to confirm the 
presence of a hydrogen-rich atmosphere on this planet.  Instead, follow-up
observations that detect the spectral signature of the planetary atmosphere 
will be necessary to confirm its chemical composition.  Fortunately, because 
the planet orbits a relatively bright and 
nearby M-star, GJ 1214b lends itself to these types of follow-up 
observations.  For a hydrogen-dominated atmosphere on GJ 1214b, the expected
transmission signature is on the level of 0.1\% relative to the host star, and
the predicted emission signature (secondary eclipse depth) is also on the order
of 0.1\%.  Hydrogen-depleted atmospheres produce comparable secondary eclipse
depths, but the transmission signature of an H-poor atmosphere is an order of
magnitude smaller -- only 0.01\%.  Current space-based instrumentation will
likely be able to detect the transmission spectrum of GJ 1214b, if the planet
does indeed posses a hydrogen-rich atmosphere.  However, under the assumption
of no strong non-LTE emission or upper-atmosphere temperature inversions, the 
next generation
of instruments aboard JWST will be needed to measure the emission spectrum
of this planet (and its transmission spectrum if the planet is depleted in 
hydrogen).  

GJ 1214b is currently the only known super-Earth that is
accessible for space-based follow-up to determine the planet's atmospheric
composition.  The only other known transiting super-Earth at this time, 
CoRoT-7b \citep{leg09} is a smaller planet orbiting a larger and hotter star.  
Even if this planet did have a significant
atmosphere, its signature would be too small to detect against the background
of its host star.  The same fate will befall the super-Earths discovered by
the Kepler mission \citep{bor04, bas05}.  Since most of the stars that Kepler 
is targeting are distant (and therefore faint) solar-type stars, the prospects
for further characterization of the planets detected around these stars are 
not strong.  One way to improve the chances of finding transiting 
super-Earths that are ideal for follow-up is by searching for planets around 
the brightest stars in the sky, such as has been proposed for all-sky missions
like the Transiting Exoplanet Survey Satellite (TESS) \citep{ric09}.  
The other method is to
search for transiting planets orbiting the smallest stars using networks of
ground-based telescopes such as MEarth 
\citep{nut07}, which is exactly how GJ 1214b was discovered.  

GJ 1214b may indeed represent a new class of planets -- those which fall in the
intermediate range between gas or ice giants, and terrestrial planets.  As more
transiting super-Earths are discovered, and the mass-radius diagram begins to
fill in for these smaller planets, it will become clearer whether there is a 
sharp transition or a smooth continuum between the two types of planets.  For
now GJ 1214b remains an interesting and unique object that is ideally suited 
for the type of follow-up observations that will reveal more information about
the nature of this planet -- both its interior and its atmosphere.  

\acknowledgements 
This work was performed in part under contract with the California Institute of
Technology funded by NASA through the Sagan Fellowship Program.  JJF
thanks the Spitzer Theory Program.


\end{document}